\documentclass[review, 3p,10pt, twocolumn]{elsarticle}

\usepackage{amssymb}
 \usepackage{amsmath}
 \usepackage{hyperref}

 \usepackage[ruled,vlined,english,linesnumbered]{algorithm2e}

\newtheorem{prop}{Proposition} 

\newtheorem{theorem}[prop]{Theorem}

  \newcommand{\tn}{\mathcal{T}}

\newcommand{\lca}{\textsc{lca}} 
\newcommand{\lsa}{\textsc{lsa}} 
\newcommand{\cluster}{\mathcal{C}} 

\newcommand{\lmap}{\theta} 

  \newcommand{\cT}{\mathcal{T}}
  \newcommand{\cB}{\mathcal{B}}

\newcommand{\pf}{\noindent{\em Proof. }}
\newcommand{\epf}{\hfill\hbox{\rule{6pt}{6pt}}}

\begin{document}

\begin{frontmatter}



\title{A cubic-time algorithm for computing the trinet
distance between level-1 networks}




\author[uea]{Vincent Moulton}
\ead{vincent.moulton@cmp.uea.ac.uk}

\author[uea]{James Oldman}
\ead{J.Oldman@uea.ac.uk}

\author[uea]{Taoyang Wu \corref{cor1}}
\ead{taoyang.wu@gmail.com}

\address[uea]{
School of Computing Sciences,
University of East Anglia,  Norwich,
NR4 7TJ, UK}

\cortext[cor1]{Corresponding author}

\begin{abstract}
In evolutionary biology, phylogenetic
networks are constructed to represent the evolution of species in
which reticulate events are thought to have occurred, such 
as recombination and hybridization. 
It is therefore useful to have efficiently computable metrics
with which to systematically compare such networks.
Through developing an optimal algorithm to enumerate all trinets displayed by a level-1 network (a type of network that is slightly more general than 
an evolutionary tree), here we propose a cubic-time algorithm to compute the 
trinet distance between two level-1 networks.
Employing simulations, we also present a 
comparison between the trinet metric and 
the so-called Robinson-Foulds phylogenetic
network metric restricted to level-1 networks.
The algorithms described in this paper 
have been implemented in JAVA and are
freely available at
(\url{https://www.uea.ac.uk/computing/TriLoNet}).
\end{abstract}


\end{frontmatter}

\section{Introduction}
\label{section:introduction}

Various types of phylogenetic networks 
have been introduced
to explicitly represent the reticulate evolutionary history
of organisms such as viruses and bacteria
in which processes such as recombination and lateral gene transfer
occur~\cite{HRS10}. 
Essentially, such networks are binary, directed acyclic graphs 
with a single root, whose 
leaves correspond to the organisms or species in question.
Here we focus on level-1 networks, a type of phylogenetic network that
is slightly
more general than an evolutionary tree, and closely 
related to so-called galled-trees (see, e.g.~\cite{cardona2011comparison}).
Level-1 networks are characterized by the property that 
any two cycles within them are disjoint (see the next section for 
a formal definition and Fig.~\ref{fig:network} for an example). 
Due to the 
availability of practical algorithms for their 
construction~\cite{lev1athan,trilonet}, level-1 
networks have attracted much attention in recent years~(see, 
e.g.~\cite{cardona2011comparison,wang2001perfect,jansson2006algorithms,gus14}) and they have been used to, for example, represent the evolution of the fungus {\em Fusarium graminearum}~\cite{HRS10}, and that of 
HBV~\cite{trilonet}.

A key challenge for phylogenetic networks is to 
quantify the incongruence between 
two networks which represent
competing evolutionary histories for a given dataset.
Such pairs can arise, for example, when different 
networks are inferred using different
methods or construction (see e.g.~\cite{huson2011survey}  
for an overview of network building methods). 
In consequence, various metrics have been developed for comparing 
phylogenetic networks  
(cf. Chapter 6 in~\cite{HRS10} for an overview).
Ideally, such a metric should be efficient to compute
since it may need to
be repeatedly computed (for example, in simulations
such as the ones that we present later in this paper).
Moreover, it is useful if the diameter 
can be derived for the metric (i.e. the maximum 
value for the metric taken 
over all pairs of all possible 
networks) so that distances can be normalized.

Here we develop an efficient cubic-time algorithm to 
compute the trinet distance between two level-1 networks,  
that is, the number of trinets (i.e., networks on three taxa) 
displayed by one but not both networks.
We also give the diameter of this metric. 
The trinet metric was introduced in~\cite{hm12} and 
used in~\cite{trilonet} to compare the performance of  
network inference algorithms.
Note that the trinet distance is closely related to 
the triplet distance, which is the number of 3-leaved trees exhibited by one but not both networks (see, e.g. \cite{jansson2014computing}). 
However, in contrast to the trinet metric,
the triplet metric is not proper in that 
there exist pairs of distinct level-1 
networks whose triplet distance is zero. 
In addition to the trinet metric, other proper metrics that 
can be used for comparing level-1 networks
include the tripartition metric~\cite{moret2004phylogenetic}, 
the path-multiplicity metric~\cite{cardona2009comparison}, 
the NNI metric~\cite{huber15space}, 
and the Robinson-Foulds metric~\cite{cardona2011comparison}. 
Among these metrics, only the NNI metric was specifically 
defined for level-1 networks, while the others were 
introduced for more general classes of networks 
and can be restricted to level-1 networks to give proper level-1 metrics.
However, establishing the diameters for these other metrics 
on level-1 metrics appears to be a challenging problem,
although in this paper we shall derive the 
diameter for the restricted Robinson-Foulds metric.

In the next section we introduce some basic notation 
and state the main result: an optimal algorithm to enumerate the trinets displayed by a level-1 network and a cubic-time algorithm to compute the trinet distance between two level-1 networks (Theorem~\ref{thm:main}).   
In Section~\ref{sec:theory} we present some 
structural results concerning level-1 networks
which we then use to prove the main result in Section~\ref{sec:algorithm}. 
 In Section~\ref{sec:experiment}
we present a comparative study
between the trinet and the Robinson-Foulds 
metrics, in which we compute some empirical 
distributions for randomly generated level-1 networks.
We conclude in Section~\ref{sec:discussion} 
with a discussion of some future directions.

\section{Preliminaries}
\label{sec:pre}

\begin{figure}[t]
\centering
\includegraphics[scale=0.7]{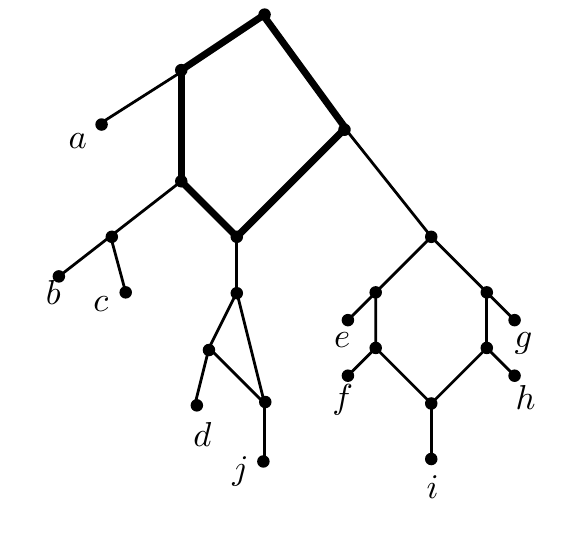}
\caption{A level-1 phylogenetic network with leaf set $X=\{a,b,\dots, j\}$ 
containing a cycle of length five, highlighted in bold. Here we use the convention that all
arcs are directed away from the root vertex which
is at the top of the network. } 
\label{fig:network}
\end{figure}

Let~$X$ be a finite set of taxa with cardinality $n$.
A \emph{rooted phylogenetic network} (or simply a 
network) $N$ on a finite set~$X$ 
is a simple, acyclic digraph with a unique root, no degenerate 
vertices (i.e., vertices with indegree one and outdegree one), 
whose leaves are bijectively labelled by the taxa in~$X$. 
A network is \emph{binary} if all non-leaf vertices 
have indegree and outdegree at most two, 
and all vertices with indegree two  
have outdegree one. A vertex is a {\em tree vertex} if it 
has outdegree two, and a {\em reticulation} if it has indegree two. 
A network is {\em level-$k$} if the maximum number of 
reticulations contained in any of its 
biconnected components is at most $k$.
Note that a network is \emph{level-1}  if it is binary 
and all of its cycles (in its underlying graph) are disjoint~\cite{HRS10}
(see Fig.~\ref{fig:network} for an example).
All networks mentioned 
in this paper, unless stated otherwise, 
are level-1.

\begin{figure}[t]
\centering
\includegraphics[scale=0.8]{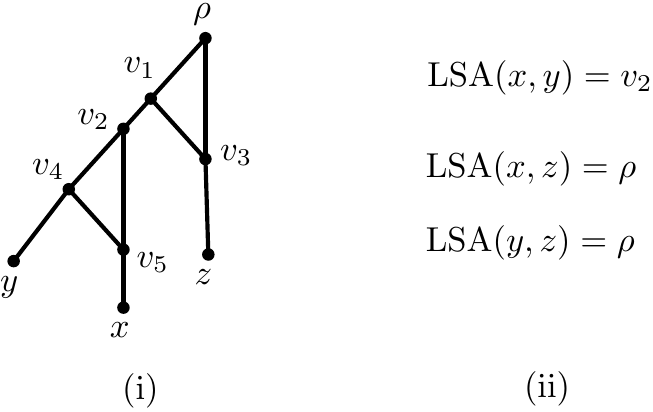}
\caption{An Example of an $\lsa$ table: (i): A level-1 phylogenetic network $N$; (ii) The $\lsa$ table of $N$.  Note that $v_4$ is the LCA of $\{x,y\}$ while we have $\lsa(x,y)=v_2$.  } 
\label{fig:lsa:table}
\end{figure}

Given a network, an arc whose removal 
disconnects the network is a {\em cut arc}. 
If a vertex $v$ is on a dipath from 
the root to a vertex $u$, then we say $u$ is 
{\em below} $v$ and $v$ is {\em above} $u$, and 
write this as $u\preceq v$ (or $u\prec v$ when $u\not =v$ holds).   
The set $\cluster(v)$ of all taxa below a
vertex $v$ is called the \emph{cluster} of $v$.  
A {\em common ancestor} of a taxon subset 
$Y$ is a vertex $v$ with $Y\subseteq \cluster(v)$. 
A {\em lowest common ancestor} (LCA) of $Y$ is a 
common ancestor of $Y$ that is not above 
any other common ancestors of~$Y$. 
A {\em stable ancestor} of $Y$ is a vertex contained in 
every dipath from the root to some taxon in $Y$. 
The \emph{lowest stable ancestor} (LSA) of~$Y$  is 
the unique vertex~$\lsa(Y)$ such that $\lsa(Y)$ is 
below every stable ancestor of~$Y$.
Note that a LCA of $Y$ is necessary 
below $\lsa(Y)$~(c.f.~\cite{fischer2010new}).
Finally, the {\em $\lsa$ table $\lmap$ of $N$} is 
the data structure 
that maps each pair of district taxa $x,y$ to $\lsa(x,y)=\lsa(\{x,y\})$ (see Fig.~\ref{fig:lsa:table} for an illustration).

\begin{figure*}[ht]
\centering
\includegraphics[scale=0.8]{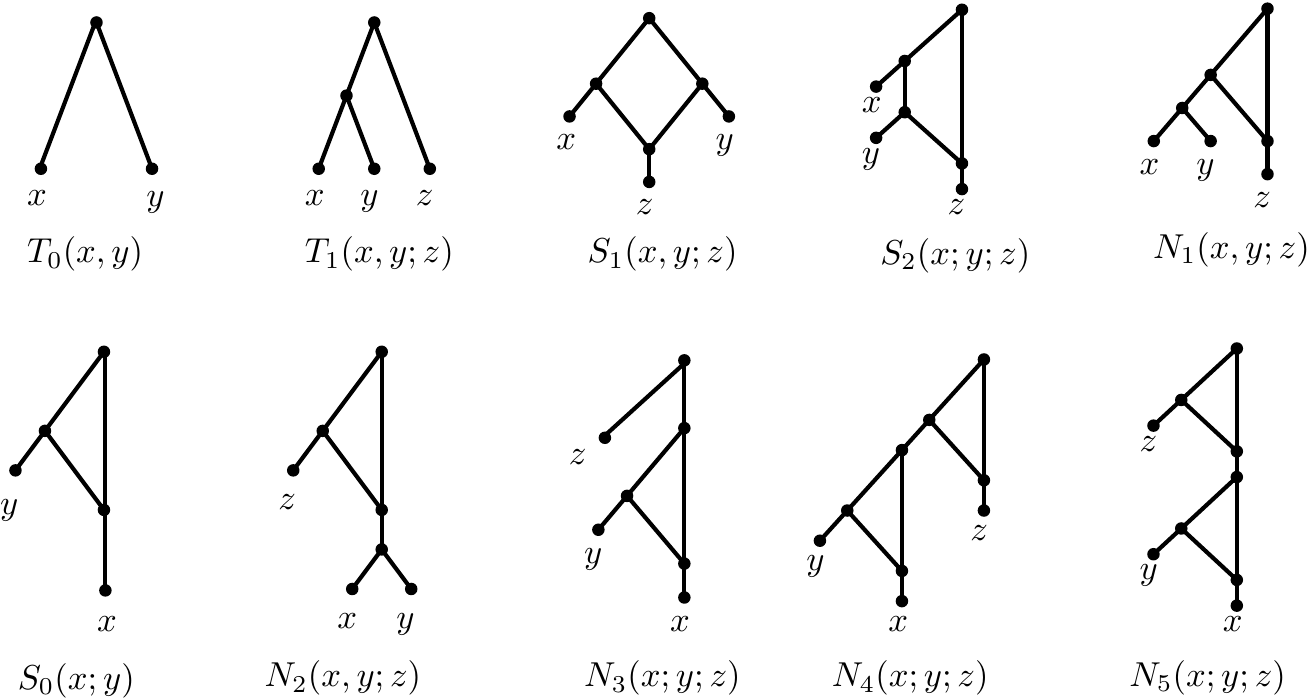}
\caption{The two types of binets and the eight types of trinets.}
\label{fig:BinetTrinets}
\end{figure*}

A \emph{binet} is a network on two taxa 
and a \emph{trinet} is a network on three taxa. 
Up to relabelling, there exist two types of binets  
and eight types of trinets~\cite{hm12}, all 
presented in Fig.~\ref{fig:BinetTrinets}.  In the following, 
we will use the notation in that figure to refer to specific 
trinets and binets.
Binets $T_0(x,y)$ and $S_0(x;y)$ are 
referred to as a {\em cherry} and a {\em reticulate cherry}, 
respectively. Note that a reticulate cherry is not symmetric, that is, $S_0(x;y)$ is distinct from  $S_0(y;x)$.

Given a network~$N$ and a taxon subset~$Y=\{y_1,\dots,y_k\}$ of $X$, 
the network $N[Y]=N[y_1,\dots,y_k]$ is the network obtained from~$N$ 
by deleting all vertices and arcs that are not on 
a dipath from $\lsa(Y)$ to some leaf in~$Y$, and repeatedly suppressing 
degree 2 vertices and replacing parallel arcs 
by single arcs until neither operation is applicable. 
Let $\cB(N)$ and $\cT(N)$ be the set of all binets and 
trinets displayed by~$N$, respectively.  It is known 
that a level-1 network $N$ is determined by 
its set $\cT(N)$ of trinets~\cite{hm12}. 

The trinet distance $d_t(N,N')$ between two 
networks $N$ and $N'$ on the set $X$ is the number of trinets 
contained in the symmetric difference $\tn(N)\triangle \tn(N')$
of the sets  $\tn(N)$ and $\tn(N')$. 
The distance $d_t$ is a metric on the set of level-1 
networks~\cite{hm12}. Moreover, 
\begin{equation}
\label{eq:tn:diameter}
d_t(N,N') \leq 2{n \choose 3},
\end{equation}
holds for any pair of networks $N,N'$ with equality holding  if, for example, 
$N$ is a tree and $N'$ is a saturated level-1 network 
(that is, each non-leaf vertex is contained in a 
cycle of size three; see~\cite{huber15space}).
Hence, the diameter of $d_t$ is $2{n \choose 3}$.
We now present our main result, whose proof 
will be presented in Section~\ref{sec:algorithm}.

\begin{theorem}
\label{thm:main}
The set  $\tn(N)$ of trinets displayed by a level-1 network $N$ on $X$  
can be constructed in $O(n^3)$ time. In addition, 
the trinet-distance $d_t(N,N')$ 
between two level-1 networks $N$ and $N'$ on $X$ 
can be computed in time $O(n^3)$.
\end{theorem}

\section{Theoretical Results}
\label{sec:theory}

In this section, we present some 
structural results concerning level-1 networks.
First, note that
given a level-1 network $N$ on $X$, we have 
\begin{equation}
\label{eq:size}
|V(N)|\leq 4n-3~\mbox{and}~ |E(N)|\leq 5n-5,
\end{equation}
with equality holding if and only if $N$ is saturated. 
The proof of this fact is similar to 
that for Lemma~1 in~\cite{huber15space}, and so we omit it.

Next, we show that in a level-1 network $N$,  each taxon subset $Y$ 
of $X$
has a unique lowest common ancestor, denoted by $\lca(Y)$. 
Note that this is not true for level-2 networks 
(see, for example,~\cite[p140]{HRS10}).

\begin{prop}
\label{prop:lca}
Each taxon subset $Y$  of $X$ has a unique lowest common ancestor  
in a level-1 network $N$ on $X$.
Moreover,  either $\lsa(Y)=\lca(Y)$ holds or 
there exists a unique  dipath from $\lsa(Y)$ to $\lca(Y)$, 
which does not contain any cut arc.  
\end{prop}

\pf
We may assume that $|Y|\ge 2$ since otherwise the proposition clearly holds.
Fix a LCA $u$ of $Y$ and let $w=\lsa(Y)$. Without loss of generality, we 
may also assume that $u\prec w$ as 
otherwise $u=w$ and the proposition follows.  

We first show that there exists a cycle in $N$ containing 
both $u$ and $w$. To this end, fix a dipath $P$ from $w$ to $u$. It 
suffices to show that $P$ contains no cut arc. If this is not
the case, let $(v_1,v_2)$ be a cut arc in $P$. 
Then $v_2\prec w$ and every dipath from the root of $N$ to 
a taxon below $v_2$ must contain $v_2$. 
Together with $u\preceq v_2$ and $\cluster(u)=Y$, this implies $v_2$ is 
a stable ancestor of $Y$, a contradiction to $w=\lsa(Y)$. 
 
It remains to show that $u$ is the unique LCA of $Y$. 
If not, let $v$ be a LCA of $Y$ with $v\not =u$. Then neither $u\preceq v$ 
nor $v\preceq u$ holds.  Now an argument similar to 
that in the last paragraph shows that $w, u$ and $v$ 
belong to the same cycle $C$.  
In addition, $w$ is the highest vertex in $C$. 
Let $P_1$ and $P_2$ be the two interior disjoint dipaths in $C$ 
so that $P_1$ contains $u$ and $P_2$ contains $v$. 
Let $u_1$ be the child of $u$ contained in $P_1$ and $u_2$ 
be the other child.  Then $(u,u_2)$ is a cut arc. 
Since $u_1$ is not a common ancestor of $Y$, there 
exists a taxon $y\in Y$ with $y\preceq u_2$. 
Since $v$ is not above $u_2$ and $(u,u_2)$ is a cut arc, $v$ 
is not above $y$, a contradiction. 
\epf

By the last proposition, a pair of distinct 
taxa $x, y \in X$ in a level-1 network $N$ on $X$ have a unique LCA, 
denoted by $\lca(x,y) = \lca_N(x,y)$. 
Moreover, $\lca(x,y)$ is precisely the interior vertex $v$ 
for which one child of $v$ is above $x$ but not $y$ while 
the other child of $v$ is above $y$ but not $x$.

A {\em splitting ancestor}  of $x$ and $y$ is an interior
vertex of $N$ such that precisely one taxon from $\{x,y\}$ 
is below both of its children 
(while the other taxon is below only one of its two children). 
For instance, in $S_0(x;y)$ the root is the unique 
splitting ancestor of $x$ and $y$ while $T_0(x;y)$ contains none.

\begin{theorem}
\label{thm:binet:type}
Suppose $x, y \in X$ are distinct taxa in a level-1 network $N$ on $X$. 
Then the following three assertions are equivalent:

\noindent
{\rm (1a)}  $N[x,y]$ is a cherry;

\noindent
{\rm (1b)}  $\lca(x,y)=\lsa(x,y)$;

\noindent
{\rm (1c)}  $x$ and $y$ do not have a splitting ancestor.

\noindent
Moreover, the following three assertions are also equivalent:
 
 \noindent
{\rm (2a)}  $N[x,y]$ is a  reticulate cherry;

 \noindent
{\rm (2b)} $\lca(x,y)\prec \lsa(x,y)$;

\noindent
{\rm (2c)}  $\lsa(x,y)$ is the unique splitting ancestor of $x$ and $y$.
\end{theorem}
 
\pf
Let $w=\lsa(x,y)$.
It is easy to show ``(1a)$\Leftrightarrow$(1b)", from 
which ``(2a)$\Leftrightarrow$(2b)" follows. 

``(1b)$\Rightarrow$(1c)": Assume $w=\lca(x,y)$ 
and that  $v$ is a splitting ancestor of $x$ and $y$. 
Let $v_1$ be the child of $v$ with $w\preceq v_1$. 
Swapping $x$ and $y$ if necessary, we assume the 
other child $v_2$ is above $x$ but not $y$.  
Consider a dipath $P_1$ from the root $\rho$ of $N$ to $v$ 
and a dipath $P_2$ from $v_2$ to $x$ such that neither of 
them contains $w$. Let $P$ be the dipath constructed by 
combining $P_1$, $P_2$, and the arc $(v,v_2)$. 
Then $P$ is a dipath from $\rho$ to $x$ that 
does not contain $w$, in contradiction to $w$ being a stable ancestor.  

``(2b)$\Rightarrow$(2c)": By Proposition~\ref{prop:lca} 
there exists a unique path $P$  from $w$ to $\lca(x,y)$. 
Fix an arbitrary vertex $u$ in $N$. Then $|\{x,y\}\cap \cluster(u)|$ 
is 2 if $w\prec u$, less than 2 if $u\prec \lca(x,y)$, 
and equal to 0 if $u$ is neither above nor below $w$. 
Therefore, $P$ contains all   splitting ancestors of 
$x$ and $y$, from which it follows that $\lsa(x,y)$ is 
the unique splitting ancestor of $x$ and $y$.

Finally, we have ``(1c)$\Rightarrow$(1b)"  as its 
contrapositive follows directly from  ``(2b)$\Rightarrow$(2c)". 
Similarly, ``(2c)$\Rightarrow$(2b)" holds as its 
contrapositive follows directly from  ``(1b)$\Rightarrow$(1c)". 
\epf

\section{Algorithms}
\label{sec:algorithm}

In this section we  present an algorithm 
for extracting trinets from a network $N$ on $X$, from 
which we can also immediately
compute the trinet distance between 
pairs of networks.

\subsection{Extracting Binets}
Our first step (see Algorithm~1) is to 
construct $\cB(N)$ and the $\lsa$ table for a 
level-1 network $N$ on $X$ in time $O(n^2)$. 

\begin{algorithm}[t]
$\lmap(x,y):=\bot$ for all $x\not =y$ in $X$ and $\cB:=\emptyset$\;

Find a topological sort $\{v_1,\dots,v_m\}$ of all 
tree vertices so that $v_j\prec v_i$ implies $i<j$ 
and construct $\cluster(v)$ for every vertex $v$\;
\For{$i=1$ to $m$}{
Let $A$ and $B$ the clusters displayed by the two children of $v_i$\;
\For{$x \in A\cap B$ and $y \in A\triangle B$}{
	$\cB \leftarrow S_0(x;y)$ and $\lmap(x,y)=v_i$\;
} 
\For{$x\in A\setminus B$ and $y\in B\setminus A$}{
 \If{$\lmap(x,y)=\bot$}{
$\cB\leftarrow T_0(x,y)$ and $\lmap(x,y)=v_i$\;}
}
}
{\bf return} the set $\cB$ and the table $\lmap$.  
\caption{Constructing $\cB(N)$ and the $\lsa$ table $\lmap$ 
for a level-1 network $N$}
\end{algorithm}

Fix a taxon subset $Y=\{x_1,x_2\}$ in $X$. We shall show 
that $N[Y]$ is the only binet on $Y$ that is contained in 
the set $\cB$ constructed in the algorithm, and that $\lmap(x_1,x_2)=\lsa(Y)$. 
The first case is that $N[Y]$ is a cherry.  Then 
by Theorem~\ref{thm:binet:type}(1), $N$ does not contain a 
splitting common ancestor of $x_1$ and $x_2$, and 
hence the pair does not occur in the for loop starting with line~4. 
In addition, by the comment below Proposition~\ref{prop:lca} 
the pair $x_1$ and $x_2$ occurs once in the for loop 
starting with line~7 (when $v_i=\lca(Y)$), from 
which it follows $N[Y]$ is the only binet on $Y$ 
contained in $\cB$, and $\lmap(x_1,x_2)=\lsa(Y)$. 

Now consider the second case in which $N[Y]$ 
is a reticulate cherry. Swapping the subscripts 
if necessary, we may assume that $N[Y]=S_0(x_1;x_2)$. 
By Theorem~\ref{thm:binet:type}, $\lsa(Y)$ is the 
unique splitting  ancestor of $x_1$ and $x_2$ in $N$  
and hence the pair $x_1$ and $x_2$ will occur in the 
for loop starting with line~5 (when $v_i=\lsa(Y)$), 
referred to here as the first event. Since $x_1$ is 
below both children of $v_i$, it follows that 
in line~6 the binet $S_0(x_1;x_2)$ is added to $\cB$ 
and $\lmap(x_1,y_2)=v_i=\lsa(Y)$. 

Next, since $\lca(Y)$ is the unique LCA of $x_1$ and $x_2$, 
the pair $x_1$ and $x_2$ will also occur once in the 
for loop starting with line~7 (when $v_j=\lca(Y)$), 
referred to as the second event. Since $\lca(Y)\prec \lsa(Y)$ 
and the vertices of $N$ are topologically sorted,  the 
first event always occurs before the second one. 
So when the second event occurs, $\lmap(x_1,x_2)$ has 
already been assigned to a vertex in $N$, and hence 
line~9 will be skipped. Therefore, $N[Y]$ is the 
only binet on $Y$ in the set $\cB$,   and  $\lmap(x_1,x_2)=\lsa(Y)$. 

Finally,  Eq.~(\ref{eq:size}) implies that $N$ 
contains $O(n)$ vertices and $O(n)$ arcs, and 
hence line~2 can be computed in $O(n^2)$. Moreover,  
the analysis in the above three paragraphs implies 
that a pair of taxa is checked precisely once in line~7 
and at most once in line~5. Therefore, the running time 
of Algorithm~1 is $O(n^2)$.

\subsection{Extracting Trinets}

Our next step is to extract trinets from a network $N$ on $X$. A 
key insight that we shall use is that each of the eight trinet types has 
a unique signature in terms of the binets it contains and 
the $\lsa$ table. 

For a network $N$ on $X$ and a triple $Y:=\{x_1,x_2,x_3\}$ in ${X \choose 3}$, 
swapping the indices if necessary, we can assume 
that $\lsa(x_1,x_2)\preceq \lsa(x_2,x_3) =\lsa(x_1,x_3)$, 
and $N[x_1,x_2]$ is either $T_0(x_1,x_2)$ or $S_0(x_1;x_2)$. 
Let $t$ be the number of cherries among the three binets on $Y$. 
Then $N[Y]$ can be inferred as follows.

\noindent
{\bf Case [$t\geq 2$]:} Let $N[Y]$ 
be $T_1(x_1,x_2;x_3)$ if $t=3$, and $N_3(x_1;x_2;x_3)$ otherwise.

\noindent
{\bf Case [$t=1$]:}  If $\lmap(x_1,x_2)=\lmap(x_1,x_3)$, 
let $N[Y]$ be $S_1(x_1,x_2;x_3)$. Otherwise if $x_3\preceq \lsa(x_1,x_2)$, 
then $N[Y]=S_2(x_1;x_2;x_3)$. Finally, let $N[Y]$ 
be $N_2(x_1,x_2;x_3)$ if $\cB$ 
contains $S_0(x_1;x_3)$, and $N_1(x_1,x_2;x_3)$ otherwise. 

\noindent
{\bf Case [$t=0$]:}   Let $N[Y]$ be $N_5(x_1;x_2;x_3)$ 
if $S_0(x_1;x_3) \in \cB$, and $N_4(x_1;x_2;x_3)$ otherwise.

Clearly, the above process can be completed in 
constant time, and hence 
the trinet set displayed by a network on $X$, as well as the trinet 
distance between two networks on $X$ can 
be computed in $O(n^3)$, from which Theorem~\ref{thm:main} follows.

\section{Experiments}
\label{sec:experiment}

To obtain some intuition concerning the empirical behaviour of 
the trinet metric, and how its behaviour compares to that of 
the Robinson-Foulds distance,
we implemented both metrics and performed experiments in 
which we computed distributions of the metrics 
for pairs of randomly generated networks. 
Note that similar experiments have been performed to understand 
properties of tree metrics~\cite{steel1993distributions} and 
RNA metrics~\cite{moulton2000metrics}.

 \begin{figure}[t]
\centering
\includegraphics[scale=0.7]{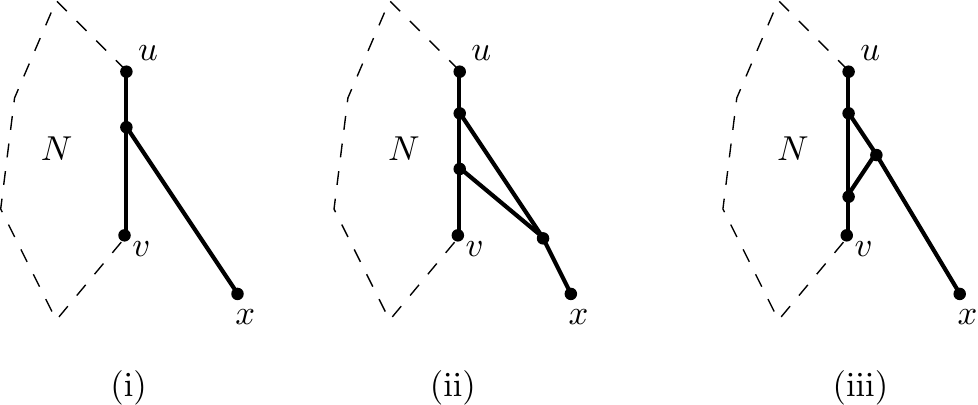}
\caption{Operations to attach a leaf to an arc $(u,v)$:
 (i): via a vertex, (ii) and (iii): via a triangle. 
Here operation (ii) and (iii) are applicable only if $(u,v)$ is not in a cycle.  } 
\label{fig:operations}
\end{figure}

We begin by recalling the Robinson-Foulds phylogenetic network metric $d_{RF}$,
which can be restricted to give a metric on level-1 networks, and 
can be regarded as a generalization of a commonly used metric on
evolutionary trees with the same name~\cite{robinson1981comparison}. 
For a pair $N, N'$ of level-1 networks on $X$ the
distance $d_{RF}(N,N')$ is defined to be
the size of the multiset that is the symmetric difference of the two 
multisets of the clusters induced by $N$ 
and $N'$~\cite{cardona2011comparison}. 
Since the root and leaf vertices of $N$ and $N'$ both induce 
identical clusters, it follows that 
$$
d_{RF}(N,N')\leq |V(N)|+|V(N')|-2-2n.
$$
Using Eq.~(\ref{eq:size}), it is 
straight-forward to check that the last inequality implies that
\begin{equation}
\label{eq:rf}
d_{RF}(N,N')\leq 6n-10
\end{equation}
holds for arbitrary pairs $N, N'$,
with equality holding in case, for example, $N$ and $N'$ are both saturated
networks that are obtained by replacing each interior vertex 
in two distinct trees whose only common clusters 
are $X$ and the singletons with 
a cycle of size three. It follows that the
diameter of $d_{RF}$ is $6n-10$.
Using Eq.~(\ref{eq:tn:diameter}) and Eq.~(\ref{eq:rf}), 
for comparison purposes in our experiments 
we normalized both the trinet and Robinson-Foulds metrics 
to take on values between $0$ and $1$.

\begin{figure*}[ht]
\centering
\includegraphics[scale=0.7]{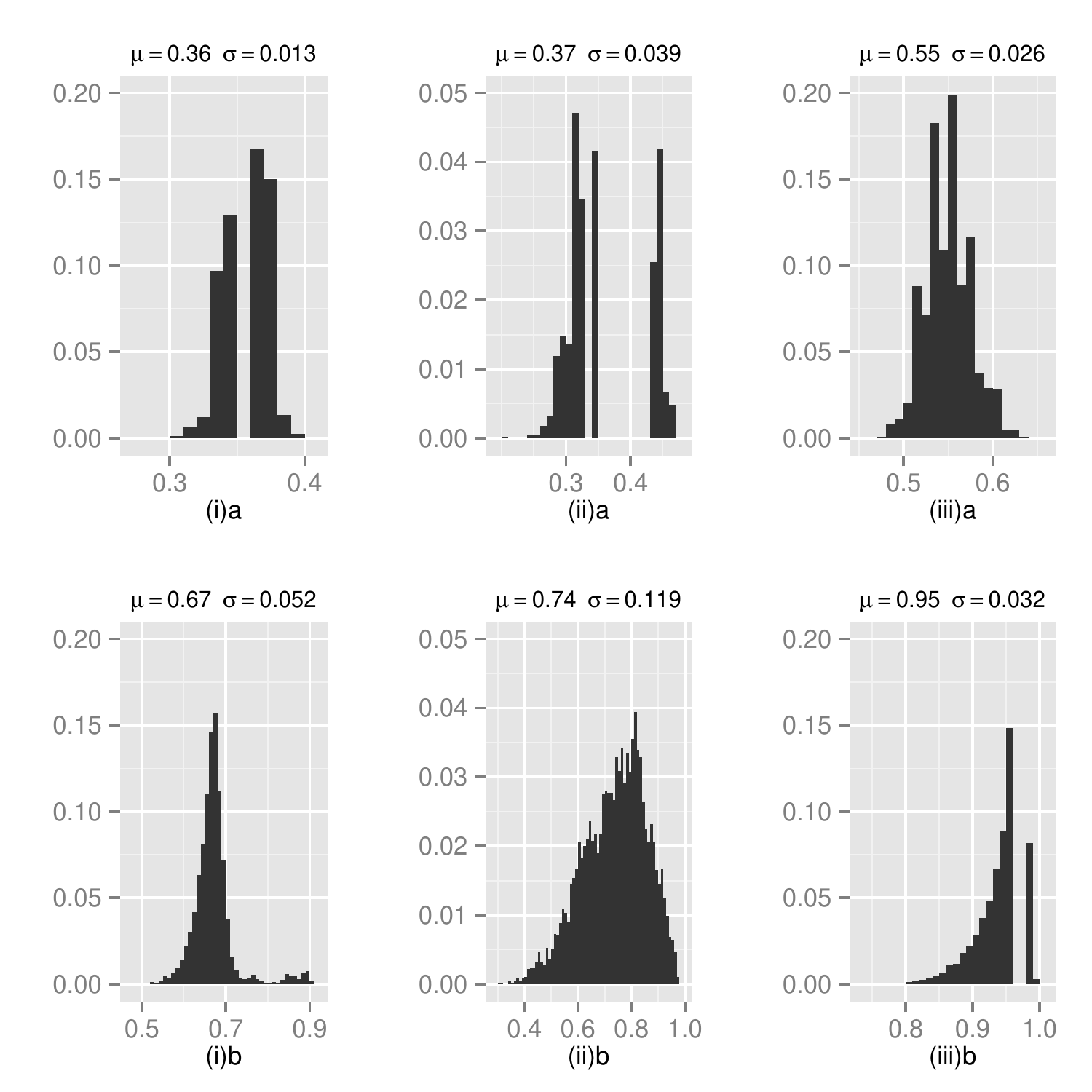}
\caption{Distributions of the normalized Robinson-Foulds  metrics (top three panels) and the trinet metrics (bottom three panels)
using bin width 0.01 on the three datasets as detailed in the text. 
The $x$-axes represent the normalized distances and the $y$-axes the proportion of network pairs with given distance. (i): Lev(1), (ii): Lev(10), and (iii): Ran. Here $\mu$ denotes the mean and $\sigma$ the standard deviation. }
\label{fig:allTrinets}
\end{figure*}
 
To perform our experiments, we generated three 
sets of random level-1 networks 
on 50 leaves in two ways as follows. 
The first two datasets,  Lev(1) and Lev(10), 
were generated under the model detailed in~\cite{lev1athan}, 
using one and ten seeds, respectively, which can be found on
 the website mentioned in the abstract.
The third dataset, Ran, was generated using the following procedure: 
starting with a random binet on two taxa, in each step 
the current network is grown by adding a taxon to a randomly 
chosen arc employing one of the three operations depicted in 
Fig.~\ref{fig:operations} until a level-1 network with the specified
leaf-set is obtained.

Distributions for the normalized trinet and Robinson-Foulds metrics
for the three datasets are presented in Fig.~\ref{fig:allTrinets}. 
For our datasets, we see that the trinet metric has a much larger range of 
values and a larger variance as compared to the Robinson-Foulds metric.
This is quite similar to the behaviour of
the Robinson-Foulds metric and quartet-distance 
evolutionary tree metrics described in 
\cite[Fig.6]{steel1993distributions}.
Note that there is a gap between the 
values presented in Fig.~(\ref{fig:allTrinets})-(ii)a, which 
could be caused by the choices of seeds in the generators 
and by the way that the metric is normalized. 

We also performed similar studies on networks with 15 and 25 leaves and 
found that as the number of leaves increases, the distribution of range 
of values got tighter for both metrics. For example, the 
standard deviations of the normalized $d_t$ and $d_{RF}$ metrics 
on the Ran datasets with 15, 25 and 50 leaves were 0.034, 0.032, 0.032,
and 0.046, 0.038, 0.026, respectively.  
We also recorded the timings for computing the two metrics. 
On a MacBook Pro computer with an i7 processor and 16 GB RAM, the 
average time for computing the trinet metric
on Lev(1), Lev(10), and Ran were 140,  145, and 231 minutes 
for the trinet metric and 16, 21, and 58 minutes for 
the Robinson-Foulds metric. Thus, as anticipated, the
Robinson-Foulds metric appears to be
somewhat faster to compute in practice.

\section{Discussion}
\label{sec:discussion}

We have presented an algorithm which allows us to 
compute the $d_t$ metric for level-1 networks, and 
demonstrated that this allows us to compute this 
metric in reasonable time for networks of up to 50 leaves. 
We have seen that although the $d_{RF}$ metric is faster to compute, 
it does not give the range of values that might be necessary to properly 
distinguish between networks. However, for certain applications $d_{RF}$ could
still serve  as a rough measure of distance suffices when timings are 
more critical. Thus, as suggested for tree and RNA 
metrics in~\cite{steel1993distributions,moulton2000metrics}, 
we do not advocate using $d_t$ or $d_{RF}$ over any other metric; 
the choice of metric will depend very much on the application.

Although our cubic-time algorithm for enumerating the trinets displayed by a level-1 network is optimal, it would be interesting to know if there is a more efficient algorithm to compute the trinet distance between two level-1 networks which does not involve listing the trinets displayed by the networks. Note that the Robinson-Foulds distance between two trees can be computed in linear time using Day's algorithm~\cite{day85} without the need to list all clusters displayed by the trees (see also~\cite{moret07} for a sublinear approximation algorithm). It may be worth exploring whether similar ideas could be exploited to more efficiently compute the trinet and the Robinson-Foulds distance between two level-1 networks.

In future work, it could be of interest to determine 
analytical formulae for the expected values and variances 
of $d_t$ and $d_{RF}$ as well as other metrics. 
Such formulae were given for different types of tree 
metrics in~\cite{steel1993distributions}.  However, as a first step
it would be probably be necessary to develop ways to generate level-1 
networks with a certain distribution (e.g. uniformly at random), 
which appears to be a challenging problem.

In addition to the two metrics studied here, as
mentioned in the introduction there are other 
proper metrics on level-1 networks (e.g. the 
tripartitions~\cite{cardona2011comparison} and
NNI~\cite{huber15space} metrics). 
However, we do not know how to normalize these 
metrics by finding their diameters. This 
is important for comparison purposes, for example, in 
the experiments that we present above. 
Therefore it would be interesting to find the diameter 
for other level-1 metrics and so that they 
can be systematically compared with $d_t$ and $d_{RF}$.

Finally, in this paper we have only considered level-1 networks, 
and it could be useful to develop efficient algorithms to 
compute trinet metrics for level-$k$ networks with $k\geq 2$,
especially
in the case $k=2$ where the trinets are known to 
determine the network~\cite{vIM12}. However,  
for networks with much higher levels this is likely to 
be challenging since, as opposed to level-1 networks, 
they are not necessarily determined 
by their trinets~\cite{huber2015much}. Therefore it might 
also be of interest to restrict to special classes of networks
(e.g. tree-child networks), where more is known 
concerning their structure~\cite{vIM12}.

\noindent
{\bf Acknowledgements}  The authors thank the anonymous referee for help suggestions on an earlier version of this manuscript. 

\bigskip
\noindent
{\bf References}


\begin{thebibliography}{10}
\expandafter\ifx\csname url\endcsname\relax
  \def\url#1{\texttt{#1}}\fi
\expandafter\ifx\csname urlprefix\endcsname\relax\def\urlprefix{URL }\fi
\expandafter\ifx\csname href\endcsname\relax
  \def\href#1#2{#2} \def\path#1{#1}\fi

\bibitem{HRS10}
D. Huson, R.~Rupp, C.~Scornavacca, Phylogenetic Networks: Concepts,
  Algorithms and Applications, Cambridge University Press, 2010.

\bibitem{cardona2011comparison}
G.~Cardona, M.~Llabres, F.~Rossello, G.~Valiente, Comparison of galled trees,
  IEEE/ACM Transactions on Computational Biology and Bioinformatics (TCBB)
  8 (2011) 410--427.

\bibitem{lev1athan}
K.~Huber, L.~van Iersel, S.~Kelk, R.~Suchecki, A practical algorithm for
  reconstructing level-1 phylogenetic networks, IEEE/ACM Transactions on
  Computational Biology and Bioinformatics 8 (2011) 635--649.

\bibitem{trilonet}
J.~Oldman, T.~Wu, L.~van Iersel, V.~Moulton, Trilonet: Piecing together small
  networks to reconstruct reticulate evolutionary histories, Molecular Biology
  and Evolution (2016) in press.

\bibitem{wang2001perfect}
L.~Wang, K.~Zhang, L.~Zhang, Perfect phylogenetic networks with recombination,
  Journal of Computational Biology 8 (2001) 69--78.

\bibitem{jansson2006algorithms}
J.~Jansson, N.~B. Nguyen, W.-K. Sung, Algorithms for combining rooted triplets
  into a galled phylogenetic network, SIAM Journal on Computing 35 (2006)
  1098--1121.

\bibitem{gus14}
D.~Gusfield, ReCombinatorics: The Algorithmics of Ancestral Recombination
  Graphs and Explicit Phylogenetic Networks, MIT Press, 2014.

\bibitem{huson2011survey}
D. Huson, C.~Scornavacca, A survey of combinatorial methods for phylogenetic
  networks, Genome biology and evolution 3 (2011) 23--35.

\bibitem{hm12}
K.~Huber, V.~Moulton, Encoding and constructing 1-nested phylogenetic networks
  with trinets, Algorithmica 616 (2012) 714--738.

\bibitem{jansson2014computing}
J.~Jansson, A.~Lingas, Computing the rooted triplet distance between galled
  trees by counting triangles, Journal of Discrete Algorithms 25 (2014) 66--78.

\bibitem{moret2004phylogenetic}
B.~M. Moret, L.~Nakhleh, T.~Warnow, C.~R. Linder, A.~Tholse, A.~Padolina,
  J.~Sun, R.~Timme, Phylogenetic networks: modeling, reconstructibility, and
  accuracy, IEEE/ACM Transactions on Computational Biology and Bioinformatics
  (TCBB) 1 (2004) 13--23.

\bibitem{cardona2009comparison}
G.~Cardona, F.~Rossello, G.~Valiente, Comparison of tree-child phylogenetic
  networks, IEEE/ACM Transactions on 
Computational Biology and Bioinformatics (TCBB)  6 (2009) 552--569.

\bibitem{huber15space}
K.~Huber, S.~Linz, V.~Moulton, T.~Wu, Spaces of phylogenetic networks from
  generalised nearest-neighbor interchange operations, Journal of Mathematical
  Biology 72 (2016) 699--725.

\bibitem{fischer2010new}
J.~Fischer, D.~H. Huson, New common ancestor problems in trees and directed
  acyclic graphs, Information processing letters 110 (2010) 331--335.

\bibitem{steel1993distributions}
M.~Steel, D.~Penny, Distributions of tree comparison metrics-some new
  results, Systematic Biology 42 (1993) 126--141.

\bibitem{moulton2000metrics}
V.~Moulton, M.~Zuker, M.~Steel, R.~Pointon, D.~Penny, Metrics on RNA secondary
  structures, Journal of Computational Biology 7 (2000) 277--292.

\bibitem{robinson1981comparison}
D.~Robinson, L.~Foulds, Comparison of phylogenetic trees, Mathematical
  biosciences 53 (1981) 131--147.
  
 \bibitem{day85}
W.~Day, Optimal algorithms for comparing trees with labeled leaves, Journal of classification 2 (1985) 7--28.

\bibitem{moret07}N.~Pattengale, E.~Gottlieb, B.~Moret, Efficiently computing the Robinson-Foulds metric, Journal of Computational Biology 14 (2007) 724--735.

\bibitem{vIM12}
L.~van Iersel, V.~Moulton, Trinets encode tree-child and level-2 phylogenetic
  networks, Journal of Mathematical Biology 68 (2014) 1707--1729.

\bibitem{huber2015much}
K.~Huber, L.~Van~Iersel, V.~Moulton, T.~Wu, How much information is needed
  to infer reticulate evolutionary histories?, Systematic biology 64 (2015)
  102--111.

\end{thebibliography}

\end{document}